\documentclass[reqno,12pt]{amsart}
\usepackage{fullpage}
\usepackage{amsfonts}
\usepackage{amssymb}

\usepackage{graphicx}
\usepackage{float}
\usepackage{caption}
\usepackage{subcaption}

\def\sech{{\rm sech}}

\def\ie/{i.e.}
\def\eg/{e.g.}

\allowdisplaybreaks[3]

\begin{document}

\title{Rogue-like waves from collision of mKdV solitons} 

\author{
Stephen C. Anco$^1$
\lowercase{and}
Jaskaran Maan$^2$
\\\\
${}^1$
D\lowercase{\scshape{epartment}} \lowercase{\scshape{of}} M\lowercase{\scshape{athematics and}} S\lowercase{\scshape{tatistics}}\\
B\lowercase{\scshape{rock}} U\lowercase{\scshape{niversity}}\\
S\lowercase{\scshape{t.}} C\lowercase{\scshape{atharines}}, C\lowercase{\scshape{anada}}
\\
${}^2$
D\lowercase{\scshape{epartment}} \lowercase{\scshape{of}} P\lowercase{\scshape{hysics}}\\
B\lowercase{\scshape{rock}} U\lowercase{\scshape{niversity}}\\
S\lowercase{\scshape{t.}} C\lowercase{\scshape{atharines}}, C\lowercase{\scshape{anada}}
}

\thanks{$^1$sanco@brocku.ca, $^2$jm22tp@brocku.ca}

\begin{abstract}
Interactions of two solitary waves having an up and a down orientation 
in the modified Korteweg-de Vries equation
are shown to produce rogue-like waves.
For waves that asymptotically vanish,
the maximum ratio between the net peak-to-trough height of the interaction profile 
and the height of the tallest incoming wave is 2.41
when the waves have approximately equal speeds,
and this ratio decreases to 2 when the speed ratio is 2.25. 
For waves that approach a non-zero constant at infinity,
the same ratio is always greater than 2.41
and reaches a maximum of 2.65 when the speed ratio of the waves is 6.32. 
\end{abstract}

\maketitle

\emph{Introduction} ---
The study of rogue waves has attracted much scientific interest in recent decades,
especially in the context of ocean waves. 
A rogue wave generally refers to a wave
that has at least twice the height of surrounding background waves 
and that may persist for only a short time or exist in isolation.

One class of analytical models of rogue waves is based on modulation instability,
such as the Peregrine soliton \cite{Per} in the nonlinear Schr\"odinger equation,
which displays both of their defining features \cite{Hen.Per.Dol}. 
There are some recent indications, however, that rogue waves in the ocean
are more likely to arise from superposition effects rather than just instability
\cite{Kno.Mal.Tay.Lib.Fed}.
In addition, it has been recognized in numerical simulations \cite{Ser.Slu} that
the largest rogue waves typically exhibit a deep trough, which contributes
significantly to the net height of the wave. 

This motivates looking at the nonlinear superposition of water wave solitons
to see when interactions of solitons may produce a \emph{rogue-like wave}
whose trough-to-peak height is at least twice the height of isolated solitary waves.
Previous studies of soliton interactions have mainly focussed on wave heights,
while ignoring troughs.

\emph{Korteweg-de Vries solitons} ---
A basic model of solitons is given by the Korteweg-de Vries (KdV) equation
\begin{equation}
  u_t + 6 uu_x + u_{xxx} =0
\end{equation}
in shallow water wave theory,
where (in dimensionless variables)  $u$ is the wave height, $t$ is time and $x$ is position.
The solitary wave solution
$u = \tfrac{1}{2} v\,\sech\big(\tfrac{1}{2}\sqrt{v}(x -vt)\big)^2$
describes an exponentially localized wave with speed $v>0$
having a height $\tfrac{1}{2}v$. 
These waves exhibit two different types of nonlinear superposition 
in which a fast, tall wave overtakes a slow, short wave \cite{Lax,Lev}. 
Their interaction is given by the 2-soliton solution 
\begin{equation}
  u = \frac{2(v_1 - v_2)(v_1\,\cosh(\tfrac{1}{2}\theta_2)^2 + v_2\,\sinh(\tfrac{1}{2}\theta_1)^2)}{\big((\sqrt{v_1} - \sqrt{v_2})\cosh(\tfrac{1}{2}(\theta_1 + \theta_2)) + (\sqrt{v_1} + \sqrt{v_2})\cosh(\tfrac{1}{2}(\theta_1 - \theta_2))\big)^2}
\end{equation}
where $\theta_i = \sqrt{v_i}(x- v_i t)$,
with $v_1>v_2>0$.
The center of mass of the two waves moves at a constant speed
and the interaction profile becomes symmetric in $x$ at $t=0$.
In terms of $r = \sqrt{v_1/v_2}$ and $v = \sqrt{v_1 v_2}$,
this profile has the form
\begin{equation}
  u =\frac{v(r^2 -1)}{2r}
  \frac{r^2\cosh(\tfrac{1}{2}\xi_2)^2 + \sinh(\tfrac{1}{2}\xi_1)^2}{\big(r\cosh(\tfrac{1}{2}\xi_1)\cosh(\tfrac{1}{2}\xi_2) - \sinh(\tfrac{1}{2}\xi_1)\sinh(\tfrac{1}{2}\xi_2)\big)^2}
\end{equation}
with $\xi_1 = \sqrt{rv}\,x$, $\xi_2 =\sqrt{v/r}\,x$.
Its shape depends only on the speed ratio $r^2>1$.
Specifically, the convexity at $x=t=0$ changes sign for $r=\sqrt{3}$.

For $r>\sqrt{3}$,
the waves undergo a merge-split,
where the two initial waves first merge into a single-peaked wave
and then split apart into two final waves with unchanged heights and speeds.
The ratio of the height of the merged wave at $x=0$
to the height of the taller wave is $1-1/r^2$.
This ratio is $2/3$ when the speed ratio has the critical value
$r_\text{crit.}=\sqrt{3}$
and increases for larger speed ratios,
reaching a maximum ratio of $1$ as the speed ratio goes to infinity.

\begin{figure}[h!]
\includegraphics[width=0.5\textwidth,trim=2cm 12cm 2cm 1cm, clip]{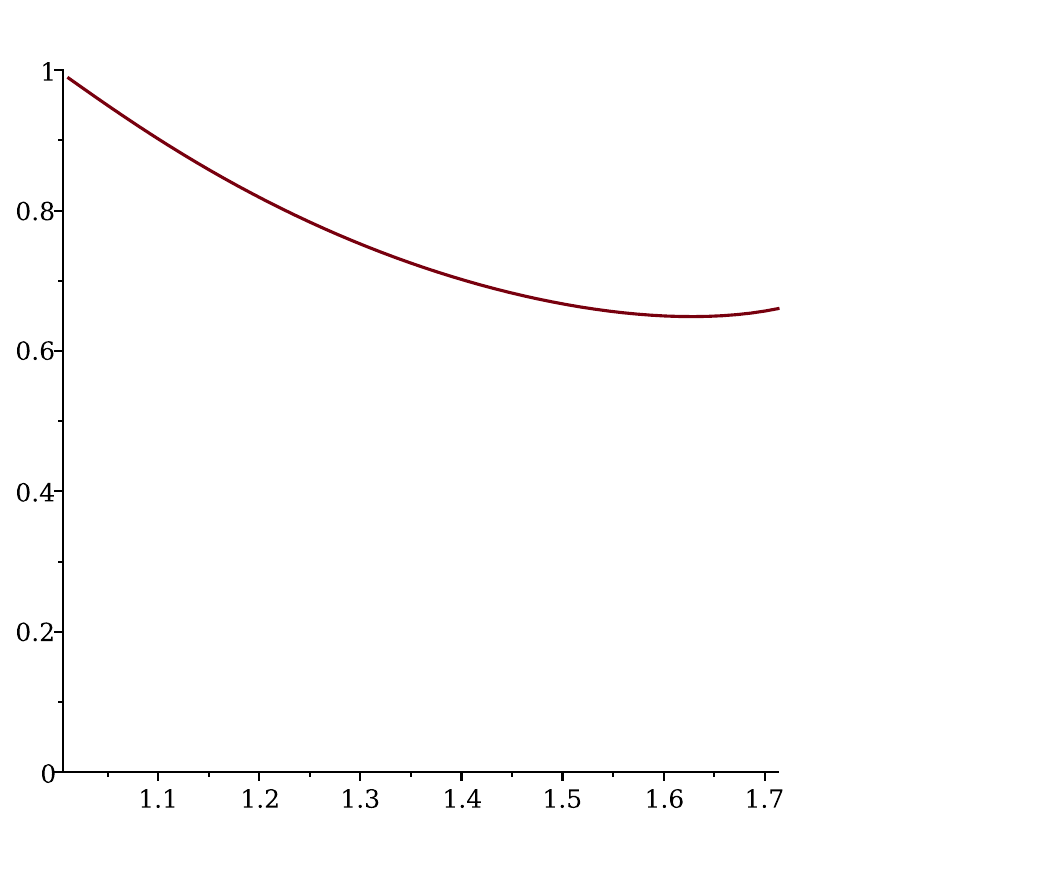}
\caption{KdV bounce-exchange: height ratio as a function of $r$}
\label{fig:kdv.sidepeak.height.ratio}
\end{figure}

For $r<\sqrt{3}$,
the waves exhibit a bounce-exchange,
where they never merge but steadily exchange properties,
and move apart with unchanged heights and speeds.
Their interaction profile at $x=0$ has a central trough and two side peaks.
The location of the side peaks $|x|=x_\text{side}$ is given by
a transcendental equation that only involves $r$ and $\zeta=x_\text{side}\sqrt{v}$.
A numerical investigation shows that the ratio of the height of the side peaks
to the height of the taller wave decreases from $2/3$ at $r_\text{crit.}$,
reaches a minimum of $0.649$ when $r=1.628$,
and increases to $1$ as $r$ approaches $1$.
See Fig.~\ref{fig:kdv.sidepeak.height.ratio}. 

Thus, no rogue-like waves are produced in interactions of two solitary waves,
as first recognized in numerical studies on KdV soliton gases \cite{Pel.Shu.Ser.Tal.Gri}. 
The situation is quite different for internal water waves.

\emph{Modified Korteweg-de Vries solitons} ---
Internal water waves in a stratified fluid \cite{Gri.Pel.Tal,Gri.Pel.Pol} 
are described by the mKdV equation 
\begin{equation}
  u_t + 6 u^2 u_x + u_{xxx} =0
\end{equation}
where (in dimensionless variables) $u$ is the wave height/depth, $t$ is time and $x$ is position.
The solitary wave solution $u = \pm \sqrt{v}\,\sech\big(\sqrt{v}(x -vt)\big)$
is similar to the KdV one, 
describing an exponentially localized wave with speed $v>0$
having a height/depth $\sqrt{v}$ in the $\pm$ cases.
Because the waves can be either up or down,
there are four different types of nonlinear superposition 
in which a fast, tall wave overtakes a slow, short wave.
These are described by the 2-soliton solution \cite{Wad}
(in the same notation used for the KdV case) 
\begin{equation}
  u = \frac{2(v_1 - v_2) (s_2\sqrt{v_2}\,\cosh(\theta_1) + s_1\sqrt{v_1}\,\cosh(\theta_2))}{(\sqrt{v_1} -\sqrt{v_2})^2\,\cosh(\theta_1+\theta_2) +(\sqrt{v_1} +\sqrt{v_2})^2\,\cosh(\theta_1-\theta_2) +4s_1 s_2 \sqrt{v_1 v_2}}
\end{equation}
where $v_1>v_2>0$, 
with $s_i=\pm1$ specifying the up/down polarity of each wave.
The center of mass of the two waves moves at a constant speed
and their interaction profile becomes symmetric in $x$ at $t=0$:
\begin{equation}
  u = \frac{2\sqrt{v}(r^2- 1)}{\sqrt{r}}\frac{s_1 r\cosh(\xi_2) + s_2\cosh(\xi_1)}{(r -1)^2\cosh(\xi_1 +\xi_2) +(r +1)^2\cosh(\xi_1-\xi_2) +4 s_1s_2 r} . 
\end{equation}

\begin{figure}[h!]
\includegraphics[width=0.45\textwidth,trim=2cm 12cm 2cm 1cm, clip]{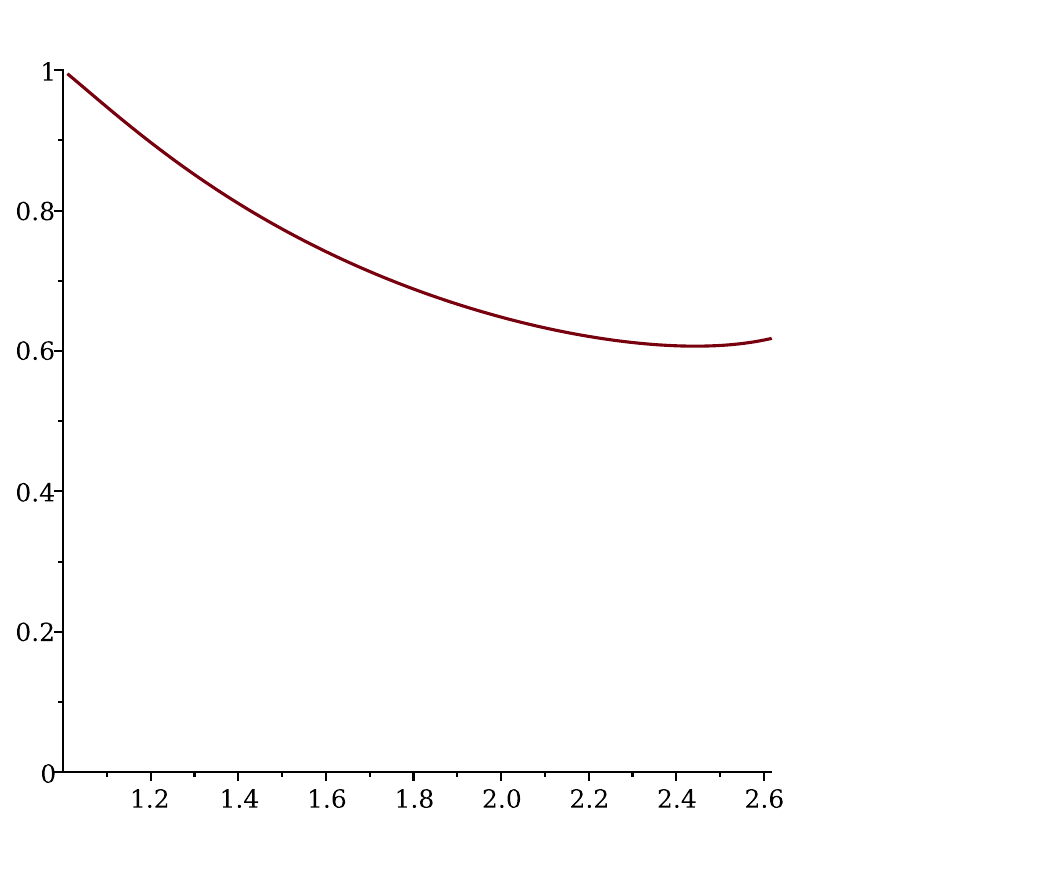}
\caption{mKdV bounce-exchange: height ratio as a function of $r$ }
\label{fig:mkdv.sidepeak.height.ratio}
\end{figure}

In interactions of two up waves, 
their profile describes either a merge-split or a bounce-exchange,
depending on whether the ratio $r=\sqrt{v_1/v_2}$ is greater or less than 
the critical value $r_\text{crit.} = \tfrac{3}{2}+\tfrac{\sqrt{5}}{2}$
for which the convexity at $x=t=0$ changes sign \cite{Anc.Nga.Wil}. 
These interactions are similar to the ones for KdV solitons.
For $r> r_\text{crit.}$, 
the ratio of the peak height to the height of the taller wave is $1-1/r$,
which equals $\tfrac{\sqrt{5}}{2} -\tfrac{1}{2}$
when the speed ratio has the critical value
and increases to $1$ as $r$ goes to infinity. 
For $r< r_\text{crit.}$,
the ratio of the side-peak height to the height of the taller wave 
decreases from $\tfrac{\sqrt{5}}{2} -\tfrac{1}{2}$ at $r_\text{crit.}$, 
has a minimum of $0.607$ when $r=2.446$,
and increases to $1$ as $r$ approaches $1$,
which is found by a numerical investigation shown in Fig.~\ref{fig:mkdv.sidepeak.height.ratio}.
The reflection $u\to -u$ gives the interaction profile of two down waves.

In contrast, interactions of a fast up wave and a slow down wave
exhibit much larger height ratios. 
Their profile describes an absorb-emit interaction
where the slow wave is first absorbed by the fast wave,
producing a symmetrical peak at $x=0$ with side troughs that form ``holes'', 
from which the fast wave then emerges and emits the slow wave behind it. 
The net height of this wave profile is given by
the height of the peak minus the depth of the troughs relative to $u=0$. 
Interactions of a fast down wave and a slow up wave are given by
the reflection $u\to -u$ of this profile.

\begin{figure}[h!]
\includegraphics[width=0.5\textwidth,trim=2cm 12cm 2cm 1cm, clip]{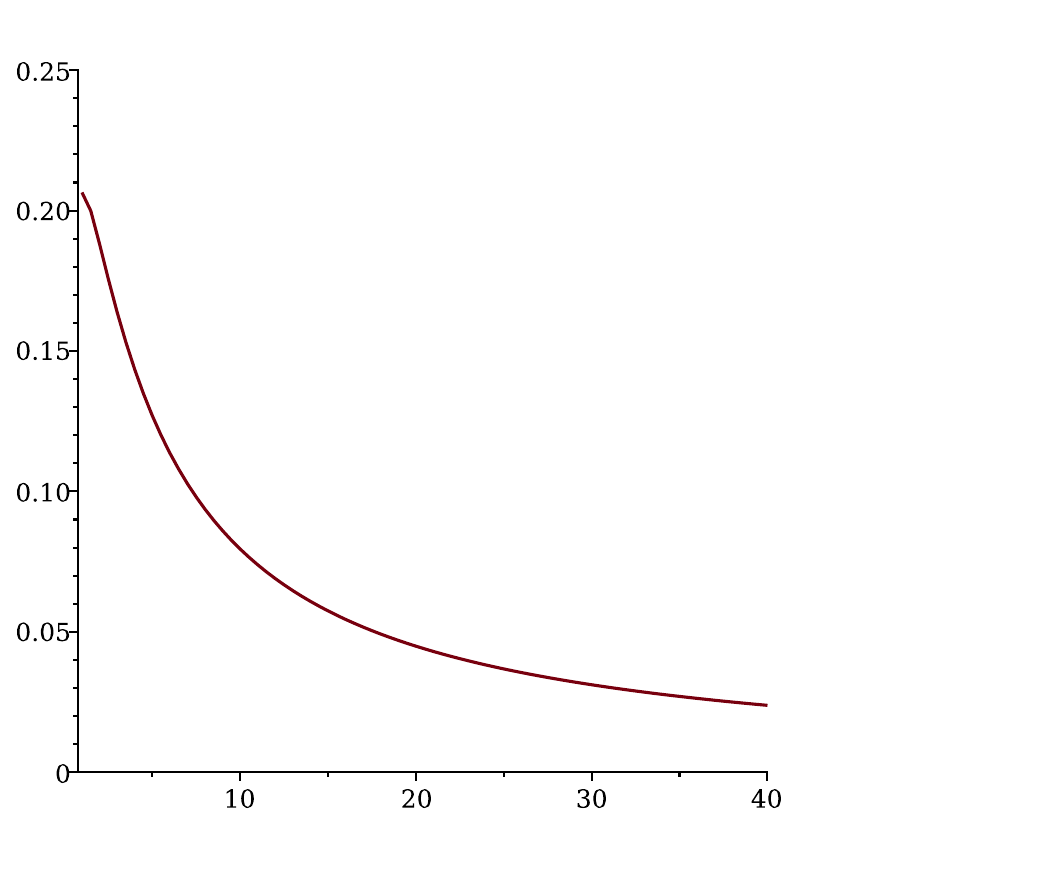}
\caption{mKdV absorb-emit: ratio of depth of trough to height of the peak as a function of $r$}
\label{fig:mkdv.troughpeak.ratio}
\end{figure}

The troughs make a significant contribution to the net height of the wave.
As shown in Fig.~\ref{fig:mkdv.troughpeak.ratio},
the depth of the trough is as large as 21\% of the height of the peak
when the speed of the up wave is close to the speed of the down wave. 

As a consequence,
the net height is significantly larger than the height of the peak. 
This observation was not taken into account in previous studies of
rogue-like waves \cite{Slu.Pel} in which the peak height was derived
but the contribution of the trough to the net height was missed. 

\begin{figure}[h!]
\includegraphics[width=0.45\textwidth,trim=2cm 12cm 2cm 1cm, clip]{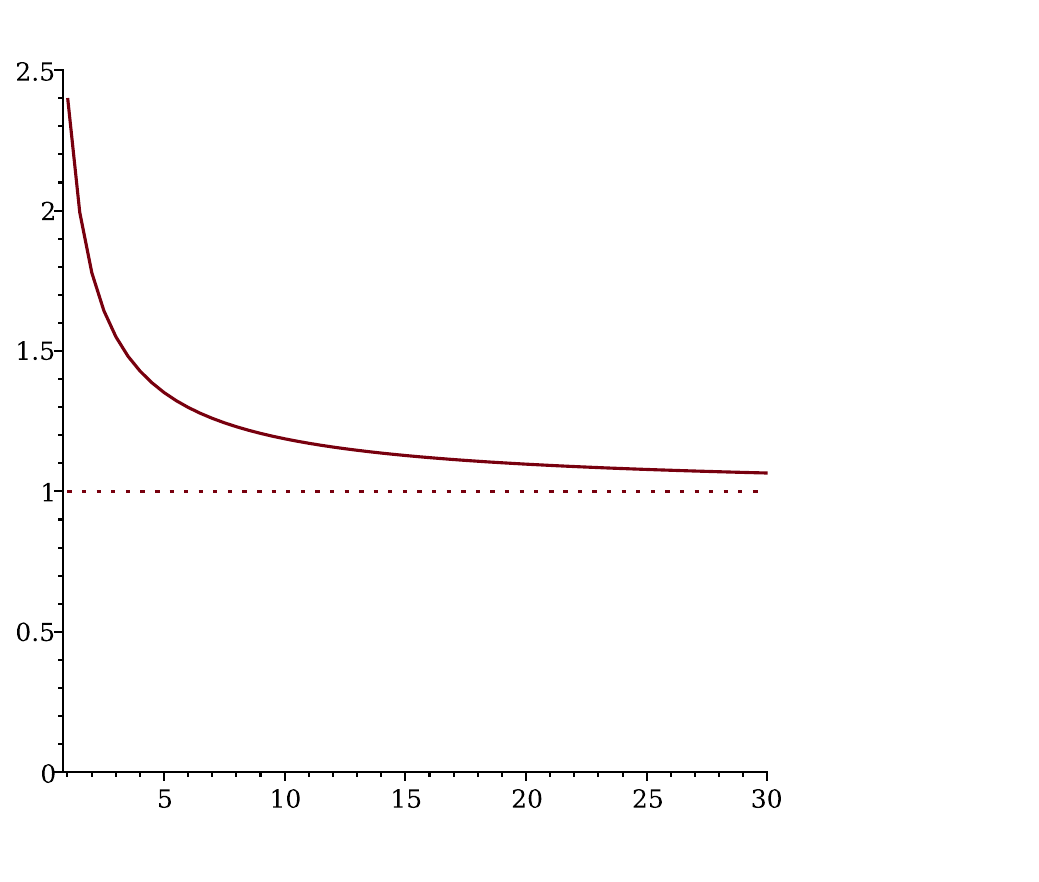}
\includegraphics[width=0.45\textwidth,trim=2cm 12cm 2cm 1cm, clip]{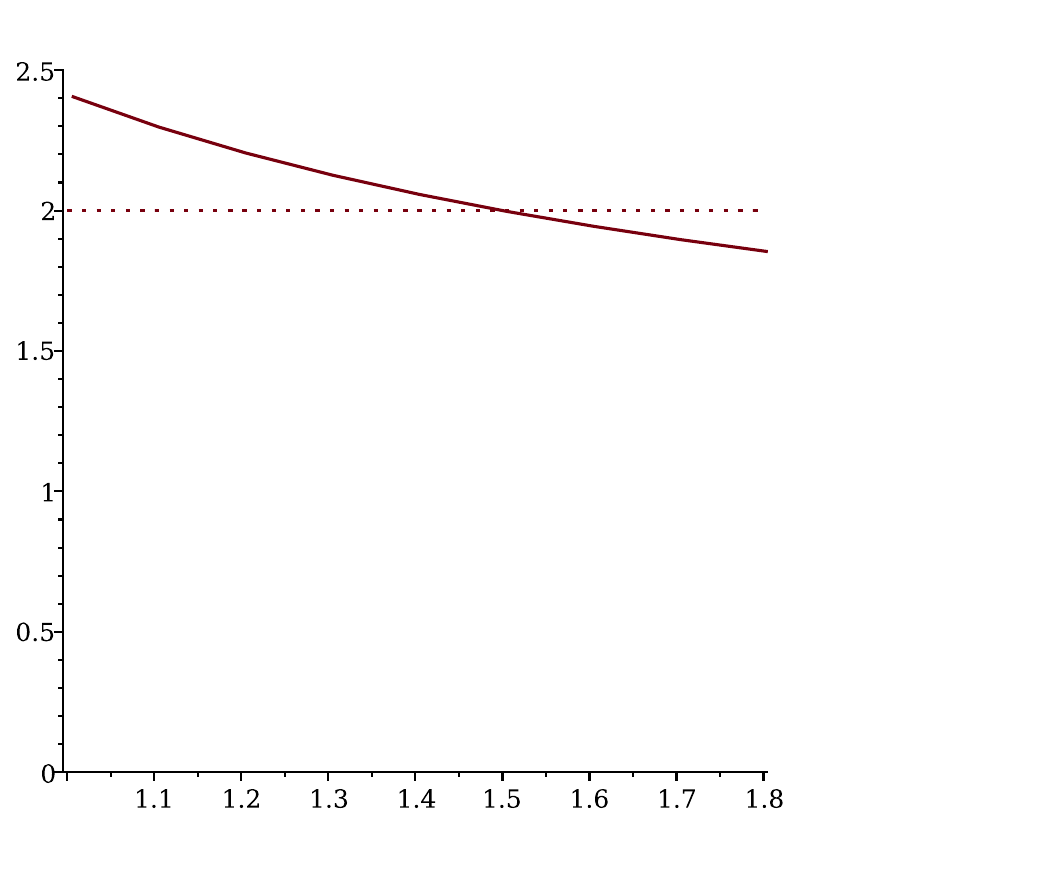}
\caption{mKdV absorb-emit: height ratio as a function of $r$}
\label{fig:mkdv.absorbemit.height.ratio}
\end{figure}

\begin{figure}[h!]
\includegraphics[width=0.45\textwidth,trim=2cm 12cm 2cm 1cm, clip]{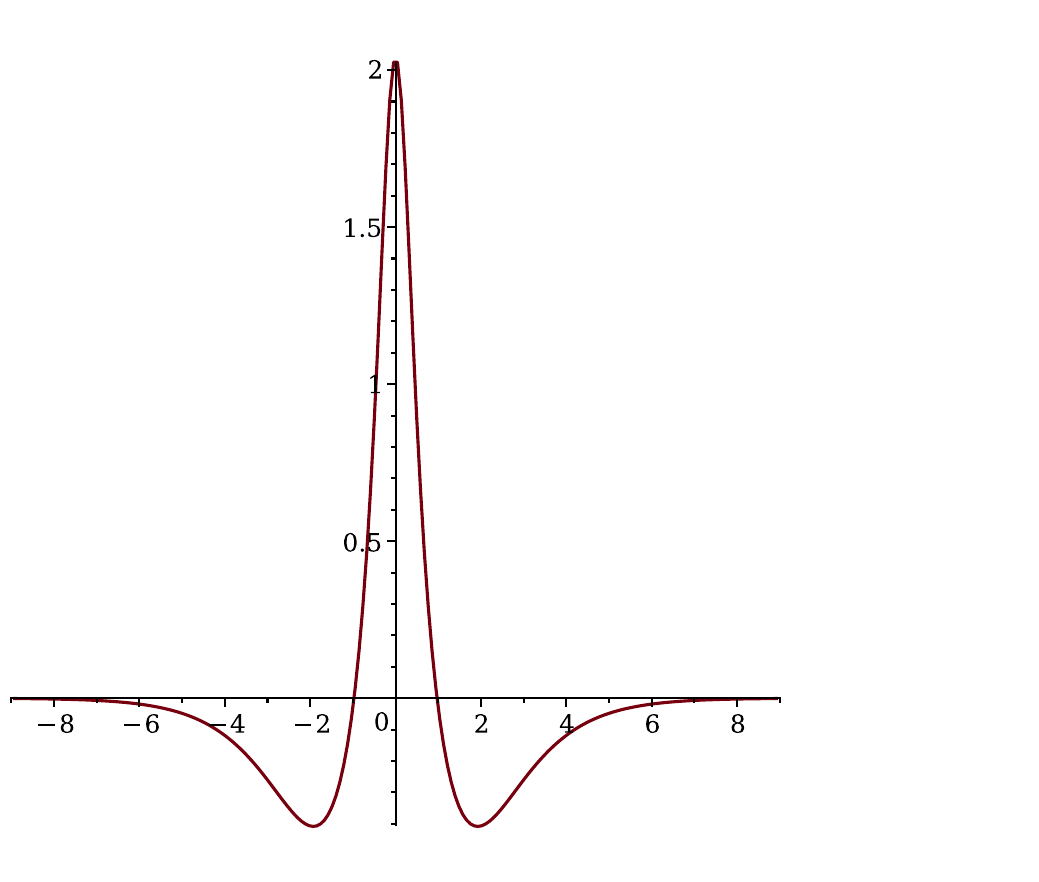}
\caption{mKdV rogue-like wave: absorb-emit profile with maximum net-height ratio}
\label{fig:mkdv.absorbemit.profile}
\end{figure}

From the numerical investigation presented in Fig.~\ref{fig:mkdv.absorbemit.height.ratio},
the ratio of the net height to the height of the up wave
approaches $2.41$ when $r\to1$;
it decreases to $2$ at $r=1.5$, and continues to decrease to $1$ as $r$ goes to infinity. 
Thus, rogue-like waves are produced in collisions of up and down waves, 
but only for a small range of relative speeds, $1< v_1/v_2 < 2.25$.
The shape of the rogue-like wave having a maximum net-height ratio is shown in
Fig.~\ref{fig:mkdv.absorbemit.profile}. 

An interesting question is what happens if the asymptotic boundary conditions of the solitons are modified.

\emph{Solitons on a non-zero background} ---
Internal waves that have a non-zero asymptotic background
arise in stratified fluids whose interface exhibits
a pedestal or offset (see e.g.\ Ref.~\cite{Gri-book}).
Given a background $b\neq0$, which can be positive or negative, 
the solitary wave solution is \cite{Jef.Kak,Mar,Anc.Nay.Rec}
$u = b +(v -6b^2)/\big(\pm\sqrt{v -2b^2}\,\cosh\big(\sqrt{v-6b^2}(x-vt)\big) + 2b\big)$,
with speed $v> 6b^2$. 
This solution takes a simpler form by parameterizing
$v=(2+ (w+1/w)^2)b^2$ in terms of $w>1$,
which yields 
\begin{equation}
  u = b\big( 1  +w_-^2/\big( \pm w_+\cosh(bw_-\xi)+ 2 \big) \big)
\end{equation}
where $w_+= w+1/w >2$, $w_-= w-1/w>0$.
Its height relative to the background is $h = u-b = \pm (w\mp1)^2 b/w$.
Notice that the relation between height and speed when the background is non-zero
differs compared to the zero-background case.
In particular, 
for any fixed speed, 
a down wave has a greater depth than the height of an up wave.

The 2-soliton solution on a non-zero background \cite{Guo.Yan}
describes four different types of nonlinear superposition 
where a fast wave overtakes a slow wave: 
\begin{equation}\label{2soliton.nzbc}
  u = b\big( 1 + A(\theta_1,\theta_2)/B(\theta_1,\theta_2) \big)
\end{equation}
in terms of $\theta_i = b w_{i-}(x - v_i t)$  and $v_i =(2 + w_{i+}^2)b^2$,
where
\begin{align}
&\begin{aligned}
A(\theta_1,\theta_2) = 
s_2 w_{1+} w_{2-}^2 \cosh(\theta_1) + s_1 w_{2+} w_{1-}^2 \cosh(\theta_2) + 2 s_1 s_2 (w_{1+}^2  - w_{2+}^2)
\end{aligned}
\\
&\begin{aligned}
 B(\theta_1,\theta_2) = & 
2 s_2 w_{1+} \cosh(\theta_1)
+ 2 s_1 w_{2+} \cosh(\theta_2)
+ \tfrac{1}{2} w_{1+}w_{2+} \Big( \frac{p_-q_+}{p_+q_-} \cosh(\theta_1 - \theta_2)
\\&\quad
+ \frac{p_+q_-}{p_-q_+} \cosh(\theta_1 + \theta_2) \Big)
+ 2 s_1 s_2 \Big( \frac{p_+p_-}{q_+q_-} +\frac{q_+q_-}{p_+p_-} \Big)
\end{aligned}
\end{align}
with $p_\pm = \sqrt{w_1w_2} \pm 1/\sqrt{w_1w_2}$, $q_\pm = \sqrt{w_1/w_2} \pm \sqrt{w_2/w_1}$.
Each wave has an up/down polarity specified by $s_i=\pm1$.
At $t=0$, their interaction profile becomes symmetric in $x$.
Similarly to the zero-background case,
this profile can be conveniently expressed in terms of
$r=\sqrt{w_1/w_2}$ and $w=\sqrt{w_1w_2}$,
where the relations $v_1>v_2$, $w_1>1$ and $w_2>1$
are equivalent to $r>1$ and $w>1$. 

For two up waves, the solution \eqref{2soliton.nzbc} 
displays either a merge-split or a bounce-exchange interaction,
determined by the convexity of the profile at $x=t=0$,
which is a function only of $r,W$.
Likewise for two down waves. 
For a fast up wave and a slow down wave,
the solution \eqref{2soliton.nzbc} instead exhibits an absorb-emit interaction,
and likewise for a fast down wave and a slow up wave. 

\begin{figure}[hb!]
\includegraphics[width=0.5\textwidth,trim=2cm 12cm 2cm 1cm, clip]{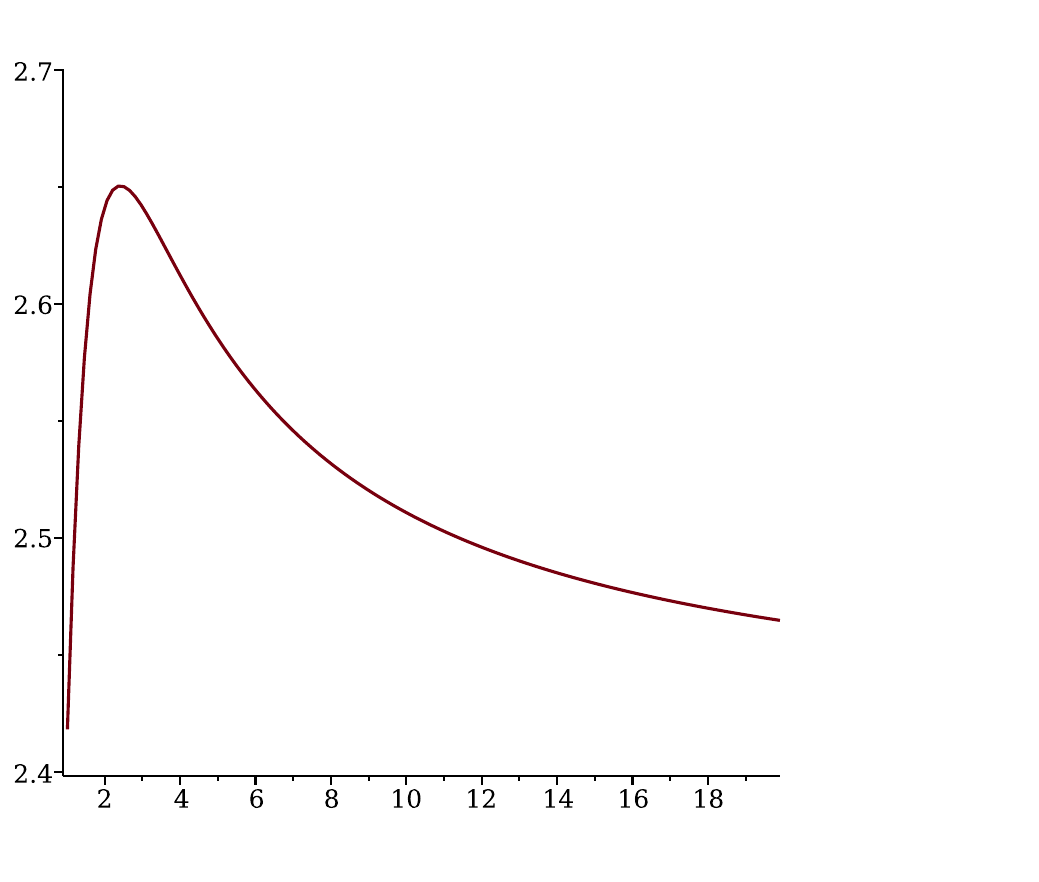}
\caption{Non-zero background mKdV absorb-emit: net-height ratio as a function of $r$}
\label{fig:nzbc.height.ratio}
\end{figure}

An analysis of the ratio between
the net height of the interaction profile 
and the height/depth of the incoming waves (relative to the background)
shows that an absorb-emit interaction produces a net height
that exceeds a ratio of 2.41
when the height of the up wave is the same as the depth of the down wave. 
This latter relation is given by $w = (r + 1)/(r - 1)$. 
Specifically,
the maximum ratio is 2.65 which occurs at $r=2.493$.
The ratio decreases to 2.41 as $r\to1$ and $r\to\infty$.
This holds independently of the size of the non-zero background $b$.
See Fig.~\ref{fig:nzbc.height.ratio}. 

\begin{figure}
\includegraphics[width=0.5\textwidth,trim=2cm 12cm 2cm 1cm, clip]{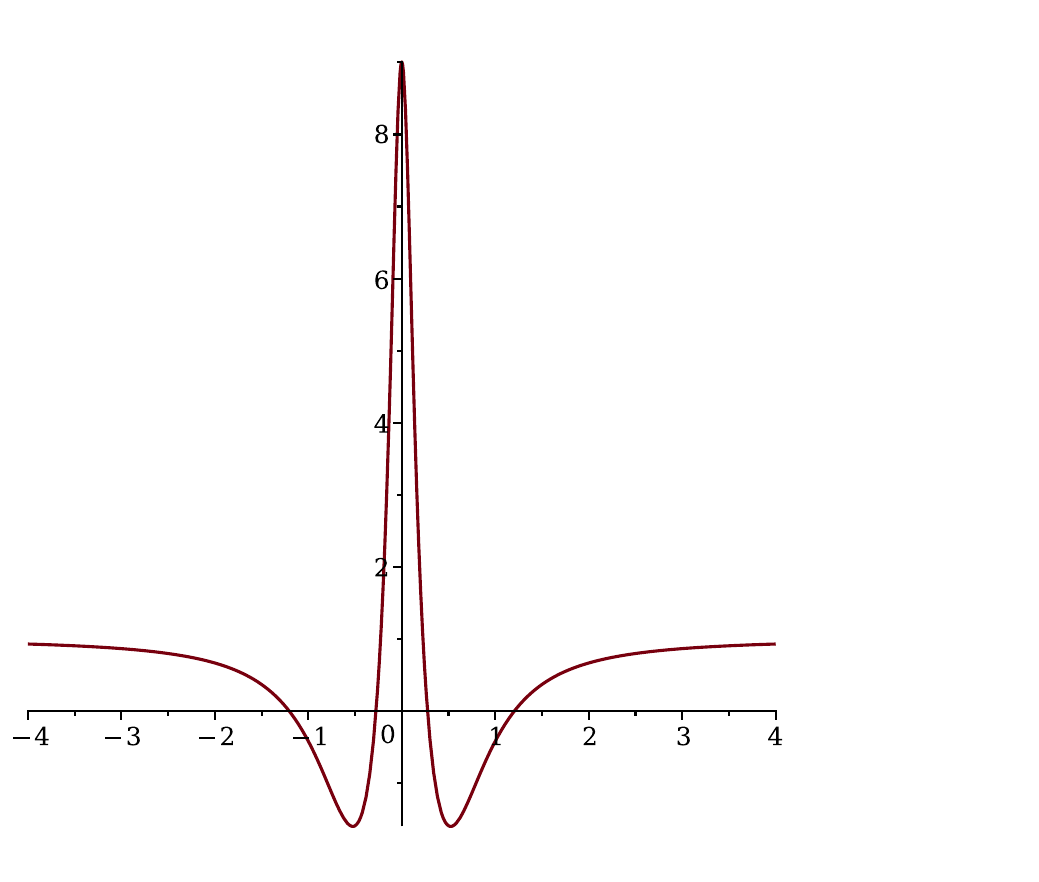}
\caption{Non-zero background mKdV rogue-like wave: absorb-emit profile with maximum net-height ratio}
\label{fig:nzbc.profile}
\end{figure}

The interaction profile exhibits a peak at $x=0$ and two side troughs, relative to $u=b$,
with the depth of the troughs being 32.5\% of the height of the peak
for the largest wave, which is quite deep.
See Fig.~\ref{fig:nzbc.profile}. 
In terms of the wave speeds,
$r = 2.493$ gives $w = 2.340$,
which yields $v_1 = 38.05 b^2$ and $v_2 = 6.02 b^2$
for the incoming up and down waves. 
Their speed ratio is $v_1/v_2 = 6.32$, which is independent of $b$. 

\emph{Conclusions} ---
Remarkably, 
rogue-like waves are always produced in mKdV collisions of an up wave and a down wave
that have equal height/depth on non-zero backgrounds. 
The ratio between the net wave height (peak to trough) and
the height/depth of the up and down waves 
ranges from 2.41 to 2.65 and is independent of the size of the background. 

Apart from describing internal water waves,
the mKdV equation also governs
anharmonic lattice waves \cite{Ono},
ion acoustic waves in plasmas \cite{Mam,Mus.Sha},
and few-cycle pulses in nonlinear optics \cite{Leb.Mih}. 
Hence, rogue-like waves will arise in these physical systems
from interactions of up and down solitary waves.

\section*{Acknowledgements}

S.C.A.\ is supported by an NSERC Discovery grant.

\end{document}